\documentclass[12pt]{article}
\usepackage{amssymb,amsmath}
\usepackage{graphicx}
\usepackage{ulem}
\usepackage[dvipsnames]{xcolor}
\begin{document} 
\begin{center} 
{\bf\Large { Puzzling Radii of Calcium Isotopes: $^{40}{\rm Ca} \rightarrow ^{44}{\rm Ca} 
\rightarrow ^{48}{\rm Ca} \rightarrow ^{52}{\rm Ca}$},
and  Duality in the Structure of $^{42}_{14}{\rm Si}_{28}$ and $^{48}_{20}{\rm Ca}_{28}$}
\end{center}
\vskip .5 cm
\begin{center}
{\bf Syed Afsar Abbas}\\ 
Centre for Theoretical Physics, JMI University, New Delhi-110025, India\\
(email: drafsarabbas@gmail.com)
\vskip .1 cm
{\bf Anisul Ain Usmani, Usuf Rahaman, Mohammad Ikram}\\
Department of Physics, Aligarh Muslim University, Aligarh-202002, India
\end{center} 
\vspace{.1in} 
\begin{center} 
{\bf Abstract}
\end{center}
In this paper we study the issue of the puzzle of the radii of calcium isotopes.
Despite an excess of eight
neutrons, strangely $^{48}{\rm Ca}$ exhibits essentially
the same charge radius as $^{40}{\rm Ca}$ does.
A fundamental microscopic description of this is still lacking.
Also strange is a peak in charge radius of calcium at N = 24.
The $^{52}{\rm Ca}$ (N = 32) nucleus,
well known to be doubly magical,
amazingly has recently been found to have a very large charge radius.
Also amazing is the property of $^{42}_{14}{\rm Si}_{28}$ which 
simultaneously appears to be both 
magical/spherical and strongly deformed as well.
We use a  Quantum Chromodynamics based model,
which treats triton as elementary entity to make up $^{42}_{14}{\rm Si}_{28}$. 
We show here how this QCD based model is able to provide a consistent physical understanding of
simultaneity of magicity/sphericity and strong deformation of a single nucleus.
This brings in an essential duality
in the structure of $^{42}_{14}{\rm Si}_{28}$ and 
subsequently also that of $^{48}_{20}{\rm Ca}_{28}$
We also provide consistent understanding of the puzzling radii of calcium isotopes.
We predict that the radius of $^{54}{\rm Ca}$ should be even bigger than that of $^{52}{\rm Ca}$; and also that the radius of
$^{60}{\rm Ca}$ should be the same as that of $^{40}{\rm Ca}$.
In addition we also show wherefrom arises the neutron E2 effective charge of $\frac{1}{2}$.



\newpage

In this paper we study the issue of the puzzle of the radii of calcium isotopes.
As stated by Garcia Ruiz {\it et al.} [1],
"Isotope shifts of stable Ca isotopes
have been extensively studied in the literature, revealing the
unusual evolution of their charge radii. Despite an excess of eight
neutrons, $^{48}{\rm Ca}$ exhibits 
the striking feature that it has essentially
the same charge radius as $^{40}{\rm Ca}$.
A fundamental explanation of these
anomalous features has been a long-standing problem for nuclear
theory for more than three decades ... but so far a microscopic description has been lacking."
Also strange is a peak in charge radius at N = 24.
The $^{52}{\rm Ca}$ (N = 32) nucleus,
through precise mass measurement and $^{54}{\rm Ca}$ (N = 34) nucleus
through study of $2^+$ have been shown to be doubly magical.
Hence we expect that this should reflect in making them more compact with small radius.
Surprisingly, however in their experiment Garcia Ruiz {\it et al.} [1]
obtained a very large radius of $^{52}{\rm Ca}$ and state that,
"The large and unexpected increase of the size of the neutron-rich 
calcium isotopes beyond N = 28 challenges the doubly magic
nature of $^{52}{\rm Ca}$  ..." Here we are able to provide a consistent 
understanding of these puzzling phenomena in a QCD based model.
We also show how a duality of the physical description of 
$^{42}{\rm Si}$ and $^{48}{\rm Ca}$ is fundamentally demanded. 

There have been conflicting claims as to the double magicity of $^{42}{\rm Si}$ [2,3,4]. 
Fridmann {\it et al.} [2,3], 
studied two-proton knockout reaction $^{44}_{16}S_{28} \rightarrow \;^{42}_{14}{\rm Si}_{28}$, 
and presented a strong empirical evidence of magicity and sphericity of $^{42}_{14}{\rm Si}_{28}$. 
However, in complete conflict with this, Bastin {\it et al.} [4] 
provided equally strong empirical evidence
showing that the N = 28 magicity had completely collapsed, 
and that $^{42}_{14}{\rm Si}_{28}$ was a well deformed nucleus.
A well developed deformation of $^{42}_{14}{\rm Si}_{28}$ was also confirmed by
Takeuchi {\it et al.} [5].
As such therefore there appears a conundrum.
So what is going on?  The majority consensus
at present however, is for no conundrum, and that
the latter experiments indicating a strongly deformed  $^{42}{\rm Si}$, have completely demolished
the earlier experimental result [6]. 

Recently we have written a paper [7], where we have demonstrated
that Fridmann experiment is actually good and correct.
They essentially explored the persistence of the exotic 
nucleus $^{42}_{14}{\rm Si}_{28}$ as a stable structure within the stable
nucleus $^{48}_{20}{\rm Ca}_{28}$, 
They showed a strong shell closure of proton number at Z = 14 which is
so dominant that it leads to extra stability, magicity and 
sphericity of $^{42}_{14}{\rm Si}_{28}$, and 
that the same is independent of the neutron magic number,
and which for this phenomenon, goes into hiding.
Thus they have found a new and novel structure of 
the exotic nucleus  $^{42}_{14}{\rm Si}_{28}$, and
which goes beyond our conventional understanding of nuclear structure which thinks 
only in terms of double-magicity with
both proton and neutron being independently and simultaneously magical.

In that paper [7], we provided a consistent understanding of 
this novel reality within a QCD based model.
This model, which has been successful in explanation of 
the halo phenomenon in exotic nuclei,
comes forward to provide the physical reason as to why the Fridmann experiment is correct. 
This QCD based model shows that it is triton, 
as elementary entity making up $^{42}_{14}{\rm Si}_{28}$, which then
provides consistency to the above amazing conclusions arising from the Fridmann experiment. 

In addition,  one of the authors (SAA) has shown [8] that 
the fusion experiment [9,10] of an incoming beam of 
halo nucleus $^{6}{\rm He}$ with the target nucleus $^{238}{\rm U}$, demonstrates that
the "core" of the halo nucleus has the structure of a tennis-ball (bubble) like nucleus,
with a "hole" at the centre
of the density distribution. This provides us with a clear-cut support
for our model of the halo nucleus [11].
This Quantum Chromodynamics based model [12,13],
had succeeded in identifying all known halo nuclei 
and made unique predictions for new and heavier halo nuclei,
and which were subsequently confirmed empirically [11,13].
It is such a potential, which is binding tritons in these neutron rich nuclei with
${^{3Z}_{\;\;Z}{\rm X}_{2Z}} = Z ({^{3}_{1}{\rm H}_{2}})$; that is,  
these nuclei are made up of Z number of tritons. 
Recently we have conducted a theoretical study within 
the ambit of the field of the RMF model structure with three
good and successful interactions. We predicted [14]  six prominent magic nuclei:
$_{\:\:8}^{24}{\rm O}_{16}$, $_{20}^{60}{\rm Ca}_{40}$,
$_{\:\:35}^{105}{\rm Br}_{70}$, $_{\:\:41}^{123}{\rm Nb}_{82}$,
$_{\:\:63}^{189}{\rm Eu}_{126}$ and $_{\:\:92}^{276}{\rm U}_{184}$.

Most significant is that the density 
distribution of $^{42}_{14}{\rm Si}_{28}$
has a hole at the centre [7].
So it looks like a tennis-ball (bubble) like nucleus. 
This is a most direct confirmation of SAA's original predictions
of 2001 [11], and discussed in detail in [8]. 
This work [7] confirms our above discussion of 
the extra stability and sphericity
of $^{42}_{14}{\rm Si}_{28}$.

Now as to the sphericity and magicity of $^{42}_{14}{\rm Si}_{28}$, 
its manifestation through only the proton number Z = 14,
and disappearance of the magic number N = 28, is extremely puzzling. 
So far we have been used to talking of sphericity and magicity
when both the proton and  neutron numbers are separately and simultaneously magical. 
However here we are being 
compelled by the empirical reality, to talk of  
sphericity and magicity of $^{42}_{14}{\rm Si}_{28}$ 
where only proton number Z = 14 shell closure is playing a role,
while the corresponding neutron number magic number
N = 28 has disappeared and gone into hiding.
This demands an understanding within our theoretical picture of nuclear physics.

Indeed, this is being provided by SAA's work of 2001 [11]
and discussed recently [8].
This QCD based model had predicted that $^{42}_{14}{\rm Si}_{28}$
has the structure of a tennis-ball/bubble
like nucleus. Also it was made up of fourteen-tritons. 
Now  triton has the  structure $^{3}_{1}{\rm H}_{2}$. 
Thus 14-tritons are a bound state 
in a potential binding these tritons as elementary entities. 
This nucleus is a extra bound state as it is closing the
triton-shell orbital $d_{\frac{5}{2}}$ at triton-number $N_t = 14$. 
This is the same as proton number Z = 14, and 
thus this is what is seen in our shell model analysis. 
As to neutrons, however, as each triton has two neutrons hidden inside
a triton (similar to the way that 2-u and 
1-d quarks are hidden inside a proton inside a nucleus),
in all 28-neutrons are hidden
inside the 14-tritons in this magical and spherical tritonic nucleus
$^{42}_{14}{\rm Si}_{28}$. Thus physically relevant is
only one magical number $N_t = 14 \sim Z = 14$. 
And it is tennis-ball/bubble like at that. 
Most importantly this model
predicts the hidden-ness of the N = 28 neutrons within the 14-tritons.
We may actually treat these 14-tritons as 14-quasi-protons,
with the same charge as protons,
but each being much heavier due to the two neutrons hidden within its guts.
Thus  $^{42}_{14}{\rm Si}_{28}$ is magical and spherical too [7].

But now Bastin {\it et al.} experiment [4] goes against the above conclusion.
It has also been confirmed recently by Takeuchi {\it et al.} [5].
So given the fact that Fridmann  {\it et al.} experiment is correct physically, 
and so is that by Bastin {\it et al.} [4] by Takeuchi {\it et al.} [5], then
there has to a duality in the physical description  of this exotic nucleus
$^{42}_{14}{\rm Si}_{28}$.
These two experiments seem to provide complementary/dual description  of
$^{42}_{14}{\rm Si}_{28}$, 
somewhat similar to the wave-particle duality necessary for understanding the structure of photon.

This possibility of duality  in the structure of 
$^{42}_{14}{\rm Si}_{28}$ gets support from the work of  
Jurado {\it et al.} [15], who in studying the masses, state in the Abstract that,
"Changes in shell structure are observed around N = 28 for P and S isotopes but not for Si.
This may be interpreted as a persistence 
of shell closure at N = 28 or as the result of very sudden 
onset in deformation in $^{42}{\rm Si}$."
Thus the two options of sphericity/magicity and strong deformation
may actually coexist simultaneously - thus 
providing support to the essential duality proposed here.

Interestingly, the concept of duality as discussed above, 
actually may be extracted from the well known and well studied
concept of "effective charge".
The polarization of closed proton shells by neutrons may be considered as inducing 
an effective charge
on the neutron for E2 transitions and quadrupole 
moments in nondeformed odd N -even Z nuclei

As emphasized by de-Shalit [16], we know that the shell model is defined 
such that an even number of protons or neutrons
couple to a zero total angular momentum. This is correct both for the
ground states of even-even nuclei as well as for the
ground states and low excited states of odd-A nuclei.
We know that as a system with zero total angular momentum in the shell model has
negligible average multipole moments, it follows that in the lowest order,  the average 
moments of odd-even nuclei are determined only by the odd group of nucleons.
Also in this model, low-lying transitions between levels of
odd-even nuclei are determined only by the odd-group of 
nucleons themselves, since the even-group for
the low excitations as well, are assumed to remain inert.

However, there is a huge problem for the above shell model, 
best expressed by de-Shalit himself [16],
"This simple model, despite its many successes, is not
adequate to describe some of the electromagnetic
properties of nuclei. As is well known, static quadrupole
moments, as well as electric quadrupole transitions, in
odd-A nuclei do not exhibit any noticeable dependence
on whether they occur in odd-Z or odd-N nuclei. The
experimental data requires that, within the framework
of the shell model, the neutron in the nucleus should be
assumed to carry an "effective charge", roughly equal to that of proton".

de-Shalit continues [16],
" ... the data on the magnetic moments seem to indicate
that the proton and the neutron retain their free
electric properties also when bound in nuclei. Thus it
seems that the "effective charge" of a neutron in the
nucleus cannot be visualized as its sharing charge with
the proton due to some exchange forces. Rather it is a
concept which is closely connected with the special
nuclear feature which is being studied,.....".
What we emphasize here is exactly this "{\bf ...it is a
concept which is closely connected with 
the special nuclear feature which is being studied,.....}".
And indeed this is pointing to an intrinsic duality 
which is forcing us to see, proton as having an extra effective charge
of $\frac{1}{2}$ in addition to its "actual" charge of unit 1, as well as
neutron as having an identical extra effective charge of
$\frac{1}{2}$ in addition to its "actual" charge of unit 0.

The standard (proton, neutron) model indeed provides a good basis for the successful
shell model of the nucleus. However this is not the whole story. 
As we have shown here, there is another
degree of freedom, the tritons, and which brings in 
a duality for a complete understanding of the nuclear phenomenon.

Now it is well known that maximum deformation in nuclei are  
expected to arise in mid-shell region between the two 
adjoining doubly magic  nuclei. A well known example is:
$^{16}_{8}{\rm O}_{8} \rightarrow \; ^{28}_{14}{\rm Si}_{14} \rightarrow \; ^{40}_{20}{\rm Ca}_{20}$,
where $^{16}_{\;\;8}{\rm O}_{8}$ and $^{40}_{20}{\rm Ca}_{20}$
are doubly magical and
$^{28}_{14}{\rm Si}_{14}$ is a well deformed nucleus.
Now as per our model, 
neutron rich nuclei are ${^{3Z}_{\; \;Z}{\rm X}_{2Z}} = Z ({^{3}_{1}{\rm H}_{2}})$; that is,  
these nuclei are made up of Z number  of tritons..
Above we have found connection between
the proton  number and the number of tritons which make up 
the nucleus ${^{3Z}_{\; \;Z}{\rm X}_{2Z}} = Z ({^{3}_{1}{\rm H}_{2}})$.
Thus magic number of Z-protons, for these triton nuclei, 
correspond to Z number of quasi-protons [7].
Hence in this chain, each
base nucleus in the chain 
 $^{16}_{\;\;8}{\rm O}_{8} \rightarrow \; ^{28}_{14}{\rm Si}_{14} 
\rightarrow \; ^{40}_{20}{\rm Ca}_{20}$,
with sufficiently extra number of neutrons should behave as follows,

$\;\;\;\;\;\;\;\;\;\;\;\;\;\;\;\;\;\;\;\;\;\;\;\;\;\;\;\;\;\;\;\;\;\; 
^{16}_{\;\;8}{\rm O}_{8} \rightarrow 
\; ^{28}_{14}{\rm Si}_{14} \rightarrow \; ^{40}_{20}{\rm Ca}_{20}$ 

$\;\;\;\;\;\;\;\;\;\;\;\;\;\;\;\;\;\;\;\;\;\;\;\;\;\;\;\;\;\;\;\;\;\;\;\;\;
\;\;\;\;\;\;\;\;\;\;\;\;\;\;\Downarrow$

\begin{equation}
^{24}_{\;\;8}{\rm O}_{16} \rightarrow \; ^{42}_{14}{\rm Si}_{28} \rightarrow \; ^{60}_{20}{\rm Ca}_{40}
\end{equation}

with $^{24}_{\;\;8}{\rm O}_{16} = 8 t \;\; ; \;^{42}_{14}{\rm Si}_{28} = 14 t 
\;\;;\;^{60}_{20}{\rm Ca}_{20} =20 t$ (t for triton).

\begin{figure}
\begin{center}
 \vspace{0.5cm}
 \includegraphics[scale=0.6]{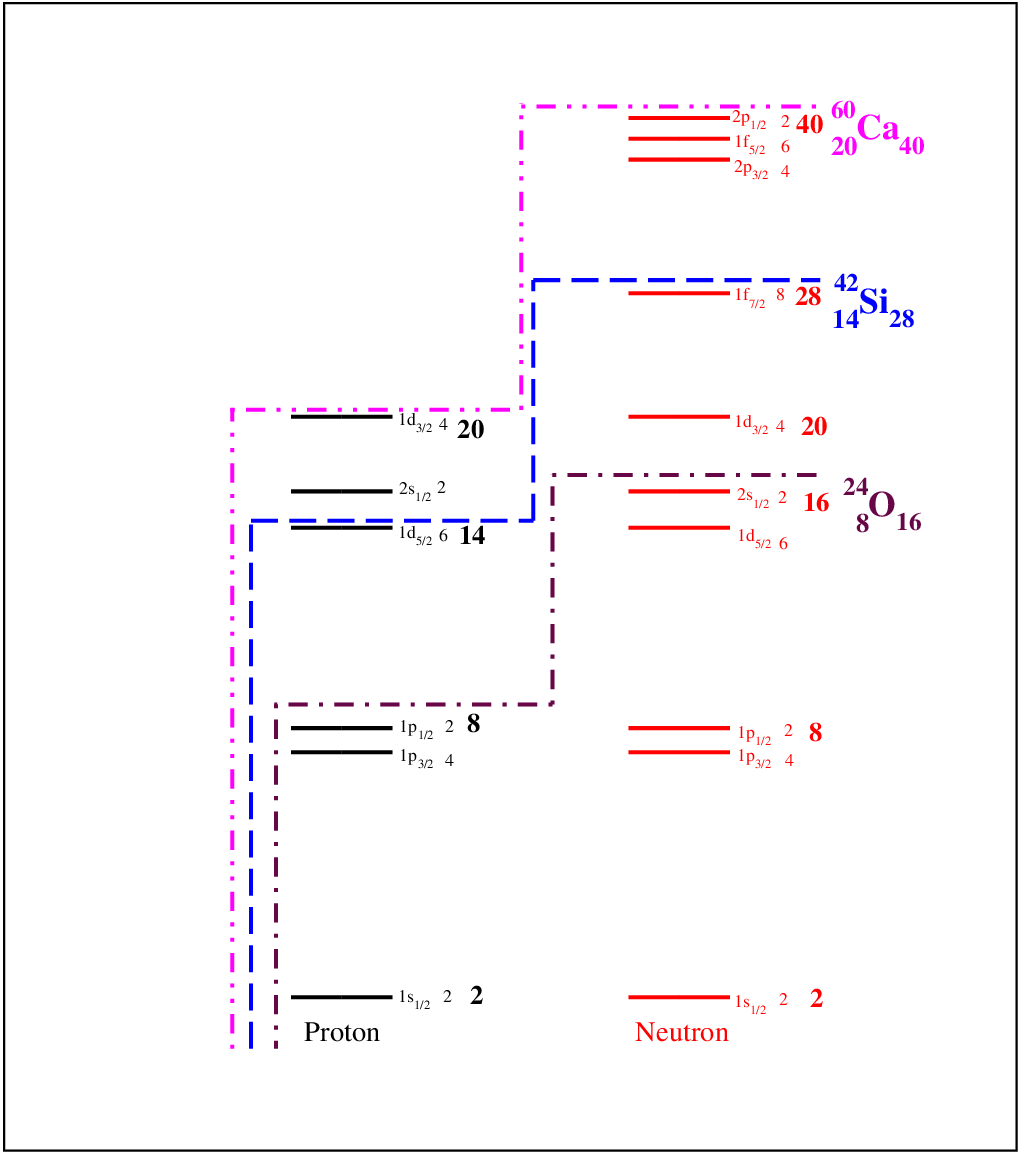}
 \caption{Shell model for proton and neutron showing how triton rich nuclei
$^{24}_{\;\;8}{\rm O}_{8} = 8 t \;\; ; \;^{42}_{14}{\rm Si}_{14} = 14 t 
\;\;;\;^{60}_{20}{\rm Ca}_{20} =20 t$
are accommodated. The level spacings here, 
however are schematic. For specific nuclei like e.g. in $^{48}{\rm Ca}$,
the proton $2s_\frac{1}{2}$ and $1d_\frac{3}{2}$ orbitals are degenerate;
and that for $^{42}_{14}{\rm Si}_{14}$, there is a significant shell gap over the filled  
proton $1d_\frac{5}{2}$ orbital [7,17].}
 \label{Fig. 1}
 \end{center}
 \end{figure}
 
Now this suggests that while 
$^{24}_{\;\;8}{\rm O}_{16}$ and $^{60}_{20}{\rm Ca}_{40}$ should be magical, the central nucleus
$^{42}_{14}{\rm Si}_{28}$ should be a well deformed nucleus.
Note that this present analysis supports
the experimental results of Bastin {\it et al.} [4] and 
of Takeuchi {\it et al.} [5], that the nucleus
$^{42}_{14}{\rm Si}_{28}$ is a deformed nucleus.

Hence, how do we expect these neutron rich nuclei:
$^{24}_{\;\;8}{\rm O}_{16} = 8 t \;\; ; \;^{42}_{14}{\rm Si}_{28} = 14 t 
\;\;;\;^{60}_{20}{\rm Ca}_{20} =20 t$,
to manifest their single-proton and single-neutron shell structure?
We show this schematically in Fig 1.
So e.g. see how $^{40}_{20}{\rm Ca}_{20} $ appears as closed 
shell doubly magical nucleus in (p,n) picture, vis-a-vis
the corresponding 20-triton structure of its 
neutron rich magical nucleus $^{60}_{20}{\rm Ca}_{40}$.

Note also how for these triton rich nuclei, 
in the  process  $^{24}_{\;\;8}{\rm O}_{16} = 8 t \rightarrow \; ^{42}_{14}{\rm Si}_{28} = 14 t$,
one needs addition of 6-tritons (i.e. 6-protons and 12-neutrons),
and how that is made possible in the shell model.
Next see how for the triton rich nuclei 
$^{42}_{14}{\rm Si}_{28} = 14 t \rightarrow \; ^{60}_{20}{\rm Ca}_{40} = 20 t$,
one again needs addition of 6-tritons (i.e. 6-protons and 12-neutrons), 
and how that is made possible in the shell model.
As shown in Fig. 1,  note that the level spacings are schematically sequential. 
However, the actual level spacings would depend upon specific nuclei, e.g. in $^{48}{\rm Ca}$  
the proton $2s_\frac{1}{2}$ and $1d_\frac{3}{2}$ 
orbitals are degenerate,
and that for  $^{42}{\rm Si}$ there is a significant shell gap over the proton
filled $1d_\frac{5}{2}$ orbital [7,17].

However as $^{42}_{14}{\rm Si}_{28}$ 
is made of $^{3}_{1}{\rm H}_{2}$, 
and  above we took it to be constituted of 14-tritons.
But note that triton has one-proton and two neutrons. Above we gave 
primacy to the number of protons and took the  number of tritons
to be equal to the number of protons. So the number of tritons may 
be treated as quasi-protons, as we indeed did [7]. 
This worked well to describe Fridmann experiment [2,3]. 
But now by ${\rm SU_{I}}(2)$ isospin-symmetry, we should not distinguish 
between protons and neutrons, as both are 
equally fundamental entities.
So neutrons should also be treatable as primary inside tritons. Hence in a conjugate/dual manner, 
let us count tritons in terms of neutrons.
So let us take the number of tritons as $N_t = 28 \sim N = 28$.
Hence it should be possible to take N = 28 of $^{42}_{14}{\rm Si}_{28}$ 
as the number of tritons. 
If this be so, 
then the corresponding proton number Z = 14
would be hidden and should go  out of contention. We propose that 
this is made possible by protons giving up their charges to the neutrons.
Thus all the charges of Z = 14 protons go over to charge the N = 28 entities. 
These neutrons therefore become quasi-neutrons, 
with each having an effective charge of $\frac{1}{2}$. 
These are counterpart of the quasi-protons [7], that
were useful to explain the success of the Fridmann experiments [2,3]. 
Hence the concept of quasi-neutrons, in a conjugate/dual manner, should be able
to explain the deformation picture of $^{42}{\rm Si}$,
as obtained by Bastin {\it et al.} [4] and 
Takeuchi {\it et al.} [5].

Next we look at Jurado {\it et al.} work 
on mass measurement of N = 28  isotones [15].
To understand odd-even staggering (OES) of
the nuclear masses, they looked at the three point indicator [18,19] defined as,

\begin{equation}
\Delta_{3} (N)=  (-1)^N  \frac{1}{2}  ( M(N-1)  + M(N+1) - 2 M(N) ) c^2
\end{equation} 

Here M(N) is mass excess of particle number.
This indicator has been found to be remarkably useful in separating 
the pairing and mean field contributions to the OES [18,19].
The indicator $\Delta_3$ for odd-N can be roughly associated with the pairing effect.
As to difference of $\Delta_3$ for adjoining odd and even 
values of N, provides information related to 
the mean field contribution in  terms of information  about the 
spacing between single particle levels.
They plotted the three point indicator for proton number Z = 14, 15, 16 and 20 in their Fig.1.
First they looked at Z = 20 case. It peaked at magic numbers N = 20 and N = 28,
as the difference of $\Delta_3$ at adjoining odd and even 
values of N is sensitive to the 
spacing between single particle levels. They also found that $\Delta_3$ for Z = 20 was 
constant between N=20 and N=28, 
this because the single particle energy is constant within a single shell
for spherical nuclei.

The case of P (Z = 15) and S (Z = 16) chain is very different from that of Ca (Z=20).
But most significant is that Si (Z = 14) behaved very similar to the case of Ca (Z=20), 
wherein between the magicity of 
Z=20 and Z=28 one observes constant single particle energy N = 20 to N = 27.
Significantly this means that exactly as in 
the case of $^{48}{\rm Ca}$ the orbital $f_{\frac{7}{2}}$ would be filled up regularly and uniformly.

Thus all the complexity of the nucleus $^{42}_{14}{\rm Si}_{28}$, unlike that of nuclei P and S, 
appears to be making it look like 
$^{42}_{14}{\rm Si}_{28}$ = $^{34}_{14}{\rm Si}_{20}$ + $\nu (f_{\frac{7}{2}})^{8}$.
It is amazing that the three point indicator is telling us that in the above,
$^{34}_{14}{\rm Si}_{20}$  as a stable core nucleus
should be doubly magical, and indeed empirically it is so. 
Then the  $f_{\frac{7}{2}}$ orbital will accommodate 
8-neutrons sequentially. This is similar to 8-neutrons 
sequentially filling up $f_{\frac{7}{2}}$ orbital
on top of doubly magical $^{40}_{20}{\rm Ca}_{20}$  
to make up  $^{48}_{20}{\rm Ca}_{28}$. 
Hence do we expect the silicon isotopes to fill up $f_{\frac{7}{2}}$
orbital in exactly similar manner as it does in the case
of calcium?

\begin{figure}
\vspace{0.5cm}
\begin{center}
\includegraphics[scale=0.5]{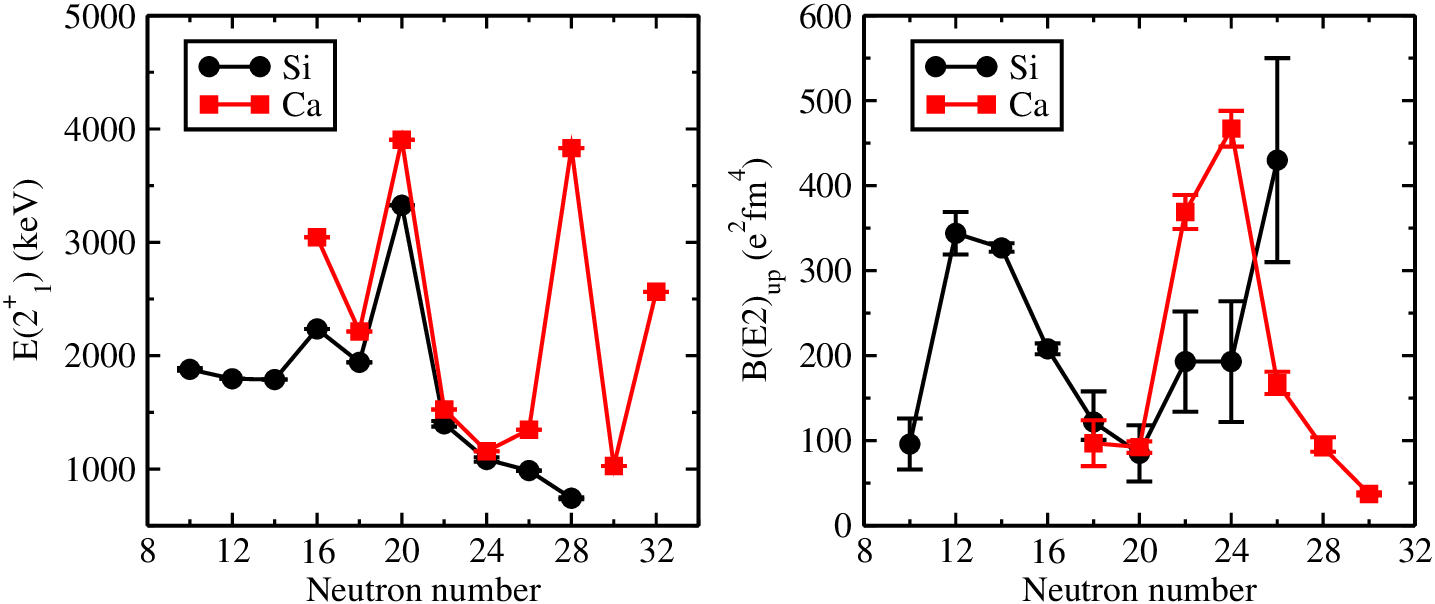}
\caption{$E(2^+$)  and $BE(2)$ experimental values [20]  for isotopes of silicon and calcium}.
\label{Fig. 2}
\end{center}
\end{figure}

We plot $E(2^+$)  and $BE(2)$ values of silicon and
calcium isotopes in Fig. 2, where experimental  values are taken from
[20] . Let us study the evolution of these in the N = 20 to 28 nuclei.
Note that magic/spherical nuclei are characterized by 
high value of $E(2^+$)  and low values of $BE(2)$;
and a deviation from this trend indicates changes in nuclear structure. 
The example of calcium isotopes  filling the $f_{\frac{7}{2}}$ 
orbital is a text-book example
of how to model multi-particle configurations of valence nucleons in an orbit [21] .
In this picture,  as there are no valence neutrons at 
the beginning of the shell at N = 20, $2^+$ energy is maximum there, 
and so also at the end of the shell (N = 28), 
where neutron excitations can no longer occur in the same shell. 
And between these two extremes, the  $2^+$ energies are almost constant. 
The corresponding B(E2) values follow a bell-shape
curve. It has  minimum values at the two extremes of the shell following
the relation $B(E2) \propto  F (F - 1)$,
where F is the fractional filling of the shell,
which here is, $ F = {\frac{(N-20)}{8}}$.

As pointed out by Sorlin and Porquet [21] , 
a breakdown of the spherical shell gap and the onset of deformation, 
would  be indicated by a deviation from this curve.
This may also happen when the configuration in a given isotopic chain for
the $2^+$ state changes from that of neutrons to that of protons.
Here we point out, that due to tritons arising as 
a new degree of freedom in our model, changes in $E^+$ and $B(E2)$
may also arise from the value of  relevant "F" itself changing.  

For the above calcium isotope case,
let us break N = 20 to 28 to N = 20 to 24 and then N = 24 to 28, 
so breaking it at the mid-point at N = 24. Then $2^+$ for calcium 
first has a sharp fall at N = 22 and then a gradual fall to N = 24.
Now for silicon identical behaviour at N = 22 and N = 24. But instead
of upturn as for calcium at N = 24, in the silicon case
the same gradual fall continues at N = 24, 26, and 28. 
As N = 28 does form a shell closure in the triton picture for silicon,
let us treat this N = 28 as what was true for calcium at N = 24,
that is at the the upturn point.  Now as our picture,
$^{24}_{\;\;8}{\rm O}_{16} = 8 t \;\; ; \;^{42}_{14}{\rm Si}_{28} = 14 t 
\;\;;\;^{60}_{20}{\rm Ca}_{40} =20 t$,
shell closure at N = 16 should connect to N=28, and the one at 
N = 28 should connect to N = 40 in these three nuclei.
Hence for silicon it is not N = 20 to N = 28 which is relevant, but N = 16 to N = 28.
{\bf And in Fig. 2 we see that this is indeed true!}.
Note that excluding the point at N = 20 for silicon, the $2^+$ 
at all the points N = 16, 18, 22, 24, 26, 28
do fall (almost) on a straight line. Thus excluding 
the N = 20 dominance point of magicity of $^{34}_{14}{\rm Si}_{20}$,
we note that what was true of calcium for N = 20 to 28,
is apparently true of silicon for N = 16 to 28. 
Hence in analogy with calcium for N = 24 to 28, we predict
that the $2^+$ for silicon would similarly 
and gradually go up in energy with a maximum at N = 40.
This is true as the lighter spherical tritonic magic nuclei are embedded in the higher spherical magic nuclei as:
$(^{60}_{20}{\rm Ca}_{40} = 20 t) \supset (^{42}_{14}{\rm Si}_{28} = 14 t) \supset  (^{24}_{8}{\rm O}_{16} = 8 t) $.
This is a unique prediction of the tritonic nature of these neutron rich nuclei.

Next, the B(E2) values of silicon isotopes. 
Clearly at N = 28 there should be a much enhanced value.
Already experimentally determined value at N=26 (Fig. 2), is attesting to this trend.
Thus we predict an even higher value of B(E2) for N = 28.
What about beyond it? Motivated by the example of calcium isotope case, we propose
the relation $B(E2) \propto  X (X - 1)$,
where  X is the fractional filling of the tritonic shell $ X = {\frac{(N-16)}{24}}$.
The peak at N = 28 is clear in this model. However near N = 20 as noted above for $E^+$,
there is a strong influence  of magicity at this number, also for B(E2) as well.
Beyond N=28 it will fall down and become zero at N=40. 
Thus note the clear-cut prediction of this model for $28 \le N \le 40 $.

Now let us study the structure of $^{48}_{20}{\rm Ca}_{28}$ .
As well known [7,17]. for this nucleus,
the proton orbital  $2s_\frac{1}{2}$ and $1d_\frac{3}{2}$  are degenerate and
are well separated from proton $1f_\frac{7}{2}$ orbital.
And as to its neutron orbital, $1f_\frac{7}{2}$ 
is well separated from $2p_\frac{3}{2}$ orbital, and
thus making it a pretty good doubly magic nucleus.
Thus as discussed above, it is a good member of the N = 20 to 28 canonical
calcium isotopic chain.
As we explained above, this is the text-book example of  
the $E^+$ and the BE(2) values for the calcium isotopes N = 20 to 28.

But there is a completely different way of looking at 
the same calcium isotope $^{48}_{20}{\rm Ca}_{28}$.
As discussed in our recent paper [7], Fridmann {\it et al.} [2,3]
essentially explored the persistence of 
the exotic nucleus $^{42}_{14}{\rm Si}_{28}$ as a stable structure within 
the stable nucleus $^{48}_{20}{\rm Ca}_{28}$, 
First Piekarewicz studied proton single particle spectrum in an  
RMF model calculation in the chain
$^{40}_{20}{\rm Ca}_{20} \rightarrow \; ^{48}_{20}{\rm Ca}_{28} \rightarrow \; ^{42}_{14}{\rm Si}_{28}$.
In Fig.3, we can see how stripping 6-protons from $^{48}{\rm Ca}$  
one gets $^{42}{\rm Si}$ . 
He showed near degeneracy of the proton orbital $1 {d}_{\frac{3}{2}} - 2 {s}_{\frac{1}{2}}$
in  $^{48}_{20}{\rm Ca}_{28}$ ;  
and the emergence of a strong  Z = 14 gap
in  $^{48}_{20}{\rm Ca}_{28}$, 
and which persisted robustly in  $^{42}_{14}{\rm Si}_{28}$.
Next, the neutron single particle spectrum behaviour was amazing.
As protons were progressively removed from the 
$1 {d}_{\frac{3}{2}} - 2 {s}_{\frac{1}{2}}$ orbitals,
$1 {f}_{\frac{7}{2}}$ neutron orbit returns 
to its parent fp-shell leading to the disappearance of the magic number N = 28.
This disappearance of the N = 28 magic number is exactly 
what Fridmann had extracted experimentally [2,3].
Note the changes necessary in the neutron orbitals to understand the above picture, in Fig.1.

Now let us see how this new neutron structure, as manifested above, 
is seen as counting the tritonic neutrons with charge 
$\frac{1}{2}$,

$\;\;\;\;\;\;\;\;\;\;\;\;\;\;\;\;\;\;\;\;\;\;\;\;\;\;\;\;\;\;\;\; 
^{48}_{20}{\rm Ca}_{28} \rightarrow \;^{42}_{14}{\rm Si}_{28} + 
\pi (1d_{\frac{3}{2}})^{4} (2s_{\frac{1}{2}})^{2} $

$\;\;\;\;\;\;\;\;\;\;\;\;\;\;\;\;\;\;\;\;\;\;\;\;\;\;\;\;\;\;\;\;\;\;\;\;\;\;\;\;\;\;\; 
\rightarrow \; ( ^{34}_{14}{\rm Si}_{20} + \nu^\prime (f_{\frac{7}{2}})^{8} ) + 
\pi (1d_{\frac{3}{2}})^{4} (2s_{\frac{1}{2}})^{2} $

$\;\;\;\;\;\;\;\;\;\;\;\;\;\;\;\;\;\;\;\;\;\;\;\;\;\;\;\;\;\;\;\;\;\;\;\;\;\;\;\;\;\;\; 
\rightarrow \; ( ^{34}_{14}{\rm Si}_{20} + \pi (1d_{\frac{3}{2}})^{4} (2s_{\frac{1}{2}})^{2}) 
+  \nu^\prime (f_{\frac{7}{2}})^{8}  $

\begin{equation}
\;\;\;\;\;\;\;\;\;\; \rightarrow \; ^{40}_{20}{\rm Ca}_{20} +  \nu^\prime (f_{\frac{7}{2}})^{8} 
\end{equation}

\begin{figure}
\begin{center}
\vspace{0.5cm}
\includegraphics[scale=0.5]{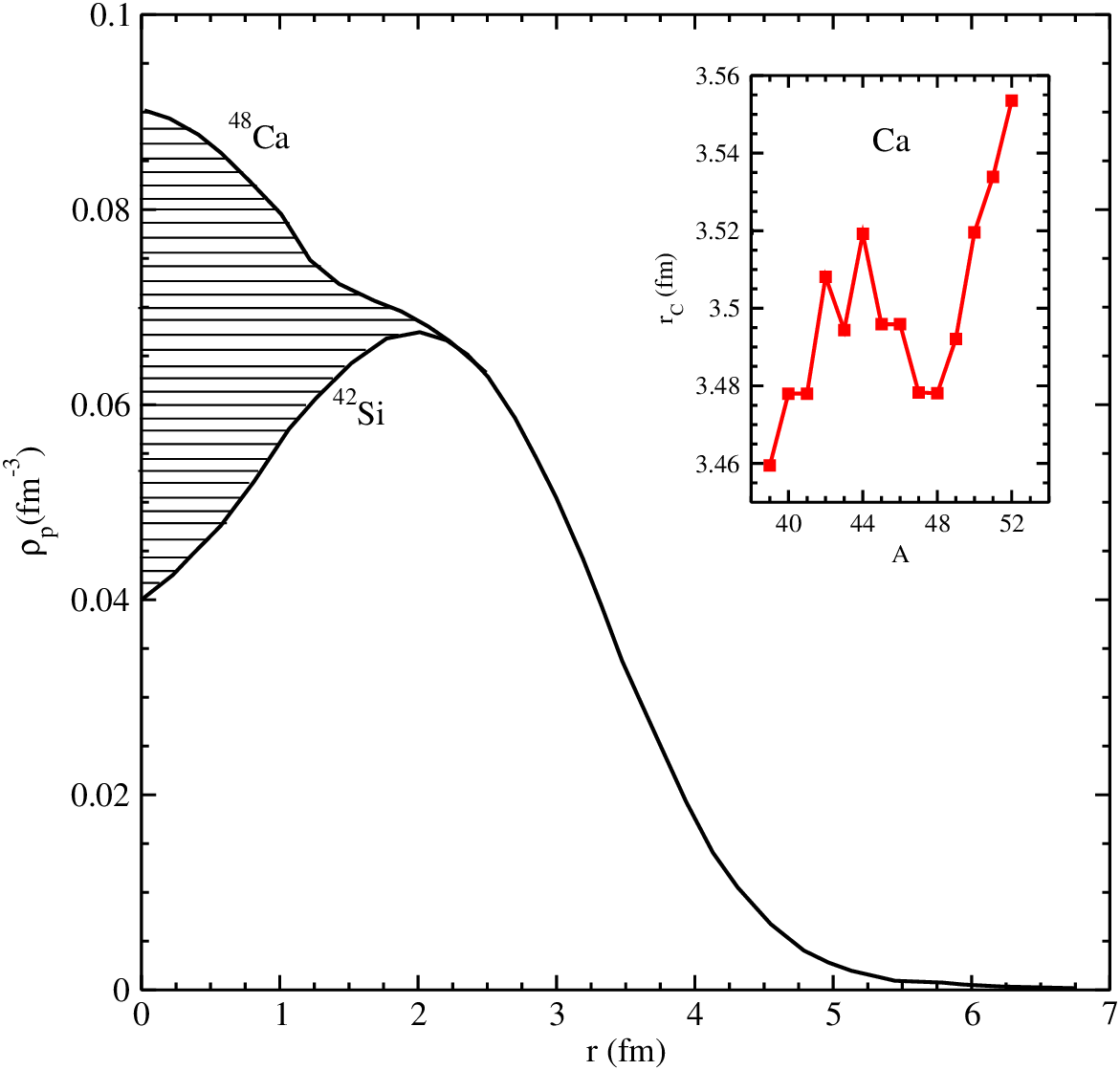}
\caption{Proton density of $^{48}_{20}{\rm Ca}_{28}$ and $^{42}_{14}{\rm Si}_{28}$ 
from [7]. Inset shows the RMS radii of calcium isotopes obtained experimentally [1]}.
\label{Fig. 3}
\end{center}
\end{figure}

The first line is supported by discussion above and also Fig. 3; 
the second line shows $^{42}_{14}{\rm Si}_{28}$
structure as discussed above, indicating that $\nu^\prime$ means 
quasi-neutrons counting the tritons as above.
The third line is rearranged to get the last line. This is justified, 
as well known, that in spite of stripping 6-protons in the chain
$^{40}_{20}{\rm Ca}_{20} \rightarrow \;^{34}_{14}{\rm Si}_{20}$, 
we go from one doubly magic nucleus to another equally very
robust doubly magic nucleus. Remember the N = 20 peak in the $2^+$ 
state of the isotopes of silicon in Fig. 2.
Thus finally we have the last line. This clearly states that 
now this $^{48}{\rm Ca}$ is made up of 8-quasi-neutrons
(and not ordinary neutrons) filling up the $f_{\frac{7}{2}}$ orbital.
And thus $^{48}{\rm Ca}$ requires 
a dual description to understand it fully.

Now where and how does this new structure of $^{48}{\rm Ca}$ 
manifest itself empirically?
We shall use this to solve the puzzle of the unexpected
radii of the calcium isotopic  chain [1].

Further, to understand the new physical reality better,  
let us try to understand the meaning of last line in eqn. (3);
$^{48}_{20}{\rm Ca}_{28}\rightarrow \; ^{40}_{20}{\rm Ca}_{20} + \nu^\prime (f_{\frac{7}{2}})^{8}$.
As $^{40}_{20}{\rm Ca}_{20}$ is 
a doubly magical and spherical nucleus,
sitting inertly in the standard (p,n) picture,
then as per the right-hand-side, $\nu^\prime (f_{\frac{7}{2}})^{8}$ 
are all quasi-neutrons with each having
an effective charge of $\frac{1}{2}$, and which arise due to
the underlying tritonic structure in this model.
Thus in the nucleus $^{48}_{20}{\rm Ca}_{28}$ , 
the core nucleus is doubly- magical and
spherical $^{40}_{20}{\rm Ca}_{20}$, 
an entity of the pure standard (p,n) shell model,
while the valence nucleons are of pure tritonic structure.
Note that this is exactly opposite in structure to the halo  nuclei, 
where as shown clearly [7,8,11,13],
the core has a pure tritonic,  hole-like and magical/spherical structure;
while the halo neutrons
arise from within a pure (p,n) shell structure.

 Hence as $^{48}_{20}Ca_{28}\rightarrow \; ^{40}_{20}Ca_{20} +  \nu^\prime (f_{\frac{7}{2}})^{8}$,
 so we predict that in this model, the structure of 41-Ca should be
 $^{41}_{20}Ca_{21}\rightarrow \; ^{40}_{20}Ca_{20} +  \nu^\prime (f_{\frac{7}{2}})$.
 Thus this predicts that for 41-Ca the
 E2 effective charge of the valence quasi-neutron is $\frac{1}{2}$, and which matches the experimental value pretty well
 [16, 22, 23]. Thus we make this amazing connection with the well known concept of E2 effective charge [16, 22, 23]
 
 Having made this connection between our quasi-neutron  charge with the E2 effective charge of $\frac{1}{2}$,
 let us try to understand how in $^{42}_{14}Si_{28}$, the  Z=14 charge is transferred to N=28.
 Here within each triton $^{3}_{1}H_{2}$, a single proton charge is transferred to a pair of neutrons.
 Thus this charge transfer should be independent of isospin.
 Given  nuclear charge as $Q = T_3 + \frac{A}{2} = \frac{Z-N}{2} + \frac{Z+N}{2}$, the charge transfer should 
 be immune to $T_3$. Thus charge of a single quasi-neutron comes from the isoscalar part $\frac{Z+N}{2}$, and hence is
 of value $\frac{1}{2}$. This automatically generates an E2 effective charge of value $\frac{1}{2}$ for proton as well.
 This is the equality of the  effective charges of proton and neutron as discussed by de-Shalit [16].
 Thus the value of the isoscalar effective charge is $\frac{1}{2} + \frac{1}{2} = 1$. Hence, this
 right away brings in, as to what is well known of the effective charge, that the isoscalar effective charge is
 of magnitude unity [16,22,23].
  
 Note that hence the valence particle's total charge is defined as:
 
 \begin{equation}
 Q_T = Q_{(p.n)\;dof} + Q_{triton\;dof}\;;\;\; Q_p = 1 + \frac{1}{2} = \frac{3}{2}, \;\; Q_n = 0 + \frac{1}{2} = \frac{1}{2}
 \end{equation}
 here dof = degree of freedom. 
 This emphasizes as to how elementary-triton-dof is as basic to the nucleus as (p,n)-dof is.

Now we are better equipped to understand the significance 
of last line of eqn. (3) -
$^{48}_{20}{\rm Ca}_{28}\rightarrow \; ^{40}_{20}{\rm Ca}_{20} + \nu^\prime (f_{\frac{7}{2}})^{8}$,
which we generalize as
N = 20 $\rightarrow$ 28, in  
$^{48}_{20}{\rm Ca}_{28}\rightarrow \; ^{40}_{20}{\rm Ca}_{20} + \nu^\prime (f_{\frac{7}{2}})^{N}$
So within $^{48}{\rm Ca}$  there sits an inert, 
magical and essentially spherical $^{40}{\rm Ca}$;
and outside it sit eight-number of quasi-neutrons
$\nu^\prime (f_{\frac{7}{2}})^{8}$, and which are of tritonic origin.

Now on to  the puzzle  of unexpected radii of the calcium isotopic chain [1] 
We display the results of their plot of radii ( their Fig 3a ) , as  inset in our Fig. 3.
Compare the calcium radii as given in the inset in Fig. 3, 
with the BE(2) values of the same nuclei in Fig. 2.
The similarity is striking, indication that the radii here 
re also behaving as a per above discussion on
BE(2) of calcium isotopes. Thus
N = 20 $\rightarrow$ 28, in  
$^{48}_{20}{\rm Ca}_{28}\rightarrow \; ^{40}_{20}{\rm Ca}_{20} + \nu^\prime (f_{\frac{7}{2}})^{N}$,
for the radii, we have minimum values at the two extremes following
the relation: radius $\propto  F (F - 1)$,
where F is the fractional filling of the shell, which here is, $ F = {\frac{(N-20)}{8}}$.
Thus, first assuming that both 40-Ca and 48-Ca have the same radii, we correctly obtain the maximum radius at 44-Ca, see Fig. 3 inset.

Now  as to why the above assumption of both 40-Ca and 48-Ca having the same radii is justified?
As we saw in eqn. (3), $^{48}_{20}Ca_{28}  \rightarrow   ^{40}_{20}Ca_{20} +  \nu^\prime (f_{\frac{7}{2}})^{8}$. Given this we
take cue from the discussion on electric charge above, we define total radius as,

\begin{equation}
N = 20 \rightarrow 28,\;in\; R_{(^{40 + N}_{20}Ca_{20+N})} = R_{(^{40}_{20}Ca_{20})} +  R_{(\nu^\prime (f_{\frac{7}{2}})^{N})}
\end{equation}

Here $R_{(^{40}_{20}Ca_{20})}  = R_{(p,n)\;dof} $  and  $R_{(\nu^\prime (f_{\frac{7}{2}})^{N})} =  R_{triton\;dof}$.
Thus the second term does not contribute to the N=20 and 28 cases. Therefore the radius of $^{48}_{20}Ca_{28}$
is the same as that of $^{40}_{20}Ca_{20}$. Hence this long standing puzzle finds a natural and consistent explanation within our model.

Next, as we saw above, $^{42}_{14}{\rm Si}_{28} \rightarrow \; ^{60}_{20}{\rm Ca}_{40}$ and with
$^{42}_{14}{\rm Si}_{28} = 14 t \;\;;\;^{60}_{20}{\rm Ca}_{40} =20 t$, we can write (see Fig. 1)

$\;\;\;\;\;\;\;\;\;\;\;\;\;\;\;\;\;\;\;\;\;
^{60}_{20}{\rm Ca}_{40} \rightarrow \;^{42}_{14}{\rm Si}_{28} + 
( 6 t \sim 6 ^{3}_{1}{\rm H}_{2} ) $

$\;\;\;\;\;\;\;\;\;\;\;\;\;\;\;\;\;\;\;\;\;\;\;\;\;\;\;\;\;\;\;\; \rightarrow \;
^{42}_{14}{\rm Si}_{28} +  ( \pi (1d_{\frac{3}{2}})^{4} (2s_{\frac{1}{2}})^{2} +  
\nu^\prime (1f_{\frac{5}{2}})^{6} (2p_{\frac{3}{2}})^{4}  (2p_{\frac{1}{2}})^{2} )$

$\;\;\;\;\;\;\;\;\;\;\;\;\;\;\;\;\;\;\;\;\;\;\;\;\;\;\;\;\;\;\;\; \rightarrow \; 
^{42}_{14}{\rm Si}_{28} +   \pi (1d_{\frac{3}{2}})^{4} (2s_{\frac{1}{2}})^{2})+  
\nu^\prime (1f_{\frac{5}{2}})^{6} (2p_{\frac{3}{2}})^{4}  (2p_{\frac{1}{2}})^{2} $ 
 
\begin{equation} 
\;\;\;\;\;\;\;\;\;\;\;\;\; \rightarrow \; ^{48}_{20}{\rm Ca}_{28} + 
+  \nu^\prime (1f_{\frac{5}{2}})^{6} (2p_{\frac{3}{2}})^{4}  (2p_{\frac{1}{2}})^{2} 
\end{equation}
where in the third line to fourth line, we have used the first line of eqn. (3).
                     
Assuming degenerate orbitals for quasi-neutrons to fill in the last line above, we have now
the mid-point at N = 34. As per
N = 28 $\rightarrow$ 40 in $ ^{60}_{20}{\rm Ca}_{40}\rightarrow 
\; ^{48}_{20}{\rm Ca}_{28} + \nu^\prime (1f_{\frac{5}{2}})^{6} 
(2p_{\frac{3}{2}})^{4}  (2p_{\frac{1}{2}})^{2}$.
for the radii, we have minimum values at the two extremes following
the relation: radius R $\propto  X (X - 1)$,
where X is the fractional filling of the shells, which here is, $ X = {\frac{(N-28)}{12}}$.
Note that it is essentially $^{40}{\rm Ca}$  which sits as robust and magical nucleus within $^{48}{\rm Ca}$  and $^{60}{\rm Ca}$ 
nuclei. 

Thus the peak in this chain would be at $^{54}{\rm Ca}$ for N = 34, 
and $^{52}{\rm Ca}$ for N = 32 would be the next highest peak  radius
in this chain. This is what Garcia Ruiz {\it et al.} found [1]. 
We have thus explained the empirical result and also here we make a unique prediction of a
peak in the radius at N = 34 for $^{54}{\rm Ca}$. Hence both 
the doubly magic calcium isotopes at N = 32 and 34 
have very large radii. 
Here we also make another significant prediction, that the radius of $^{60}{\rm Ca}$  should also be the 
same as that of $^{40}{\rm Ca}$ .
The unique role of tritons in this above model has to be appreciated. 

In summary, the nuclear description, in addition to (p,n) shell structure, needs
a tritonic degree of freedom at an equally fundamental level.
Inclusion of this new degree of freedom explains the duality existing within the structures of $^{42}_{14}Si_{28} \;and\; of\; ^{48}_{20}Ca_{28}$ nuclei. We are thus able to explain and understand the origin of the puzzling radii of the calcium isotopes.
We are also able to understand the origin of the E2 isoscalar effective charge of value 1, with the neutron and the proton 
having the same value of $\frac{1}{2}$ for their individual effective charges.

\newpage
{\bf REFERENCES:}

1. R. F. Garcia Ruiz, M. L. Bissell, K. Blaum, A. Ekstroem, N. Froemmgen, G. Hagen, M. Hammen,
K. Hebeler, J. D. Holt, G. R. Jansen, M. Kowalska, K. Kreim, W. Nazarewicz, R. Neugart,
G. Neyens, W. Nörtershäuser, T. Papenbrock, J. Papuga, A. Schwenk, J. Simonis,
K. A. Wendt and D. T. Yordanov, Nat. Phys. 12 (2016) 594

2. J. Fridmann {\it et al.}, Nature 435 (2005) 922

3. J. Fridmann {\it et al.}, Phys. Rev. C 74 (2006) 034313

4. D. Bastin {\it et al.}, Phys. Rev, Lett, 99 (2007) 022503

5. S. Takeuchi {\it et al.}, Phys. Rev. Lett. 109 (2012) 182501

6. A. Gade {\it et al.}, Phys. Rev. Lett. 122 (2019) 222501

7. S. A. Abbas, A. A. Usmani, U. Rahaman, M. Ikram, arxiv:1907.10342

8. S. A. Abbas, "Fusion of halo nucleus....", Mod. Phys. Lett. A - IN PRESS Aug. 2019;
DOI: 10.1142/S0217732319502213

9. R. Raabe {\it et al.}, Nature 431 (2004) 823

10. D. Hinde and M. Dasgupta, Nature 431 (2004) 748

11. A. Abbas, Mod. Phys. Lett. A 16 (2001) 755 

12. A. Abbas, Phys. Lett. B 167 (1986) 150; Prog. Part. Nucl. Phys., 20 (1988) 181 

13. S. A. Abbas, "Group Theory in Particle, Nuclear, 
and Hadron Physics", Taylor and Francis Group, USA, 2016 

14. A. A. Usmani, S. A. Abbas, U. Rahaman, M. Ikram and 
F. H. Bhat, Int. J. Mod. Phys. E 27 (2018) 1850060

15. B. Jurado {\it et al.}, Phys. Lett. B   649 (2007) 43

16. A. de-Shalit, Phys. Rev. 113 (1959) 547

17. J. Piekarewicz, J. Phys. G: Nucl. Part. Phys. 34 (2007) 467

18. W. Satula, J. Dobaczewski, W. Nazarewicz,Phys. Rev. Lett. 81 (1998) 3599 
 
19. J. Dobaczewski, P. Magierski, W Nazarewicz,  W. Satula, and Z. Szymanski, 
Phys. Rev. C 63 (2001) 024308

20. B. Pritychenko {\it et al.}, At. Data Nucl. Data Tables 107 (2016) 1

21. O. Sorlin and M-G. Porquet, Phys. Scr. T152 (2013) 014003

22. L. Zamick, M. Golin and S. Moszkowski, Phys. Lett. B 66 (1977) 116;
A. K. Dhar and K. H. Bhatt, Phys. Rev. C 16  (1977) 792; 
T. Minamisono et al, Z. Naturforsch. 57 a (2002) 595;
T. K. Alexander, B. Castel, I. S. Towner, Nucl. Phys. A 445 (1985) 189

23. A. Abbas and L. Zamick, Phys. Rev. C 21 (1980) 731

\end{document}